# Validating Coordination Schemes between Transmission and Distribution System Operators using a Laboratory-Based Approach

Filip Pröstl Andrén[*], Thomas I. Strasser[*], Julien Le Baut[*], Marco Rossi[†], Giacomo Viganò[†],
Giacomo Della Croce[‡], Seppo Horsmanheimo[§], Armin Ghasem Azar[¶] and Adrian Ibañez[‖]
[*]Center for Energy, Electric Energy Systems, AIT Austrian Institute of Technology, Vienna, Austria
{filip.proestl-andren, thomas.strasser, julien.lebaut}@ait.ac.at
[†]Ricerca sul Sistema Energetico (RSE), Milan, Italy, {marco.rossi, giacomo.vigano}@rse-web.it
[‡]SELTA S.p.A., Piacenza, Italy, giacomo.dellacroce@selta.com
[§]VTT Technical Research Centre of Finland, Espoo, Finland, seppo.horsmanheimo@vtt.fi
[¶]Technical University of Denmark (DTU), Lyngby, Denmark, agaz@dtu.dk
[‖]Our New Energy, Alacant, Spain, aip@ournewenergy.com

*Abstract*—The secure operation of future power systems will rely on better coordination between transmission system and distribution system operators. Increasing integration of renewables throughout the whole system is challenging the traditional operation. To tackle this problem, the SmartNet project proposes and evaluates five different coordination schemes between system operators using three benchmark scenarios from Denmark, Italy, and Spain. In the project, field tests in each of the benchmark countries are complemented with a number of laboratory validation tests, to cover scenarios that cannot be tested in field trials. This paper presents the outcome of these laboratory tests. Three tests are shown, focusing on controller validation, analysis of communication impacts, and how well price-based controls can integrate with the SmartNet coordination schemes. The results demonstrate important indications for the field tests and also show some of the limitations with the current implementations of the coordinations schemes.

## I. INTRODUCTION

The continuously integration of large quantities of Renewable Energy Sources (RES) is challenging the whole European power system, both at the transmission and the distribution level. One commonly proposed solution is ancillary services provided by connected units. For a secure operation of the network all units—including RES, flexible loads, and storage systems—should provide such services to the grid. On top of that, better coordination between the operation of transmission and distribution grids will be necessary for the future [1].

The SmartNet project proposes and evaluates five coordination schemes between Transmission System Operator (TSO) and Distribution System Operator (DSO) [2]. This work is complemented with different real-time market architectures with the aim of finding out which of the solutions would deliver the best compromise between costs and benefits for the whole system. Each coordination scheme presents different ways of organizing relationships between system operators. The main focus in the SmartNet project is to study simulation scenarios of three benchmark countries, i.e., for Italy, Denmark, and Spain. The simulations are then scaled-up to a full replica lab where the performance of real controller devices will be tested. At the same time, three demonstration projects (pilots) for testing the specific technological solutions are implemented to enable monitoring, control, and participation in ancillary services provision from flexible entities located in distribution. For this purpose, one pilot is executed in each of the aforementioned benchmark countries [2].

Although many different viewpoints and issues can be covered in the three pilots, there are still certain aspects that cannot be covered or only partially analyzed. The laboratory tests are dedicated to replicate and test some of the functionalities to be implemented within the pilots. Here, the focus of the lab tests is to anticipate some potential issues and troubles before they are implemented in the real scenarios. In addition to that, laboratory tests can add further possibilities to test new functions that cannot be tested in a pilot. For example when the current regulatory framework is blocking [3].

Having these goals in mind, the laboratory tests will focus on evaluating certain equipment that were conceived for purpose of the pilots. This is done by combining the capabilities of a laboratory test environment with the SmartNet simulator [4]. In other words, the laboratory validations will be based on Hardware-In-the-Loop (HIL) setups, avoiding relying only on software simulations [5]. This paper summarizes some of the findings from these laboratory investigations.

This paper is organized as follows: Section II presents the methodology used for the validations. Section III gives an overview of the used validation environment. In Section IV the three implemented validation cases are presented, which are followed by the conclusions in Section V.

## II. METHODOLOGY

As described above, the laboratory tests are associated with the SmartNet pilots. One of the main goals with this work is to anticipate some potential issues and troubles before they are implemented in a real scenario (i.e., the pilots). Because of this, the scenarios developed in the pilots were used as main motivation for collecting possible laboratory validation and test cases. The following main goals were planned for this work: *(i)* analyze selected equipment and elaborate suggestions on their utilization within the pilots, *(ii)* validate additional hardware components that were not directly covered in the pilots, and *(iii)* analyze how Information and Communication Technology (ICT) aspects, such as latency, packet drops, etc. affect the performance of the coordination schemes.

The focus of this paper is to see how today's components and technologies can be integrated with the future coordination scenarios developed within the SmartNet project. The used validation methodology was inspired by the ERIGrid validation approach [6] and can be described by the following steps:

1) *Collecting validation cases:* In the first phase different use cases were collected by analyzing the three pilots.
2) *Selection of test cases:* Each of the selected validation cases from the first step were analyzed in more detail and an assessment was made whether the validation case should be implemented in the laboratory or not.
3) *Test and experiment specification:* Based on the selected test cases it needs to be specified how the test should be executed and what experiments are part of the test.
4) *Experiment execution and collection of results:* The defined experiments are executed and measured results are collected. Iterations with the previous steps may be needed in case the results are not descriptive enough.

This paper mainly reports the findings from steps 3 and 4.

## III. VALIDATION ENVIRONMENT

The validation environment consist of two main parts: the SmartEST laboratory and the SmartNet simulator.

### A. The SmartEST Laboratory

The SmartEST laboratory offers an environment for testing, verification, and R&D in the field of distributed energy systems and smart grid applications. The laboratory accommodates Distributed Energy Resource (DER) components, such as inverters, storage systems, and voltage regulators/controllers. Controllable AC and DC sources allow a testing capability up to 1 MVA. Additional equipment for simulating control and communication interfaces and an environmental chamber offer extended testing capabilities [7].

Complementary to the power system components the laboratory also includes real-time Power Hardware-In-the-Loop (PHIL) simulations, which can be used to combine hardware system tests with the advantages of numerical simulations. By means of a controllable AC voltage source, simulated distribution network models can be coupled with real components. This allows a more rapid development, validation, and evaluation of control algorithms, system concepts and components for Smart Grid applications [7]. Fig. 1 shows a simplified schematic of the SmartEST laboratory.

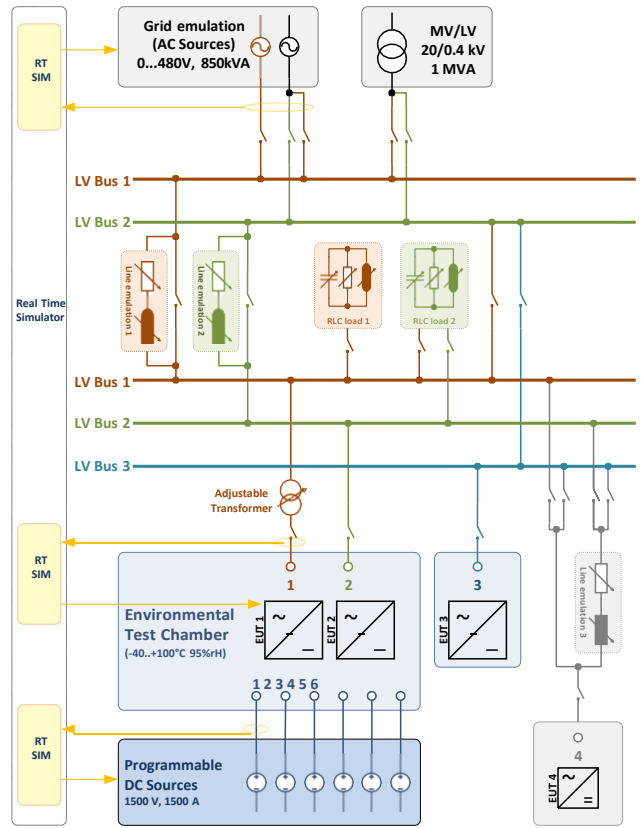

Fig. 1. Simplified schematic of the SmartEST laboratory [7].

For this work, the SmartEST infrastructure including the Low-Voltage (LV) grid as well as loads (representing the consumer behavior), energy storage systems but also distributed generators (i.e., solar Photovoltaic (PV) inverters) are used in different scenarios. On top of that, the available lab automation system will be used for test automation, control commands, and logging of measurements.

### B. The SmartNet Simulator

One of the main results of SmartNet is the simulator, capable of simulating and analyzing the different TSO-DSO coordination schemes proposed by the project. As described above, the SmartEST laboratory will be used to integrate real hardware components and controllers. This is done through a HIL setup where the components are interfaced with the SmartNet simulator. The simulator consists of three layers [4]:

- *Physical layer*: The basis of the simulator is represented by a physical simulation of the system, covering the behavior of each network (transmission and distribution) and their components with automatic controls (e.g., secondary frequency regulation).
- *Bidding and dispatching layer*: The interface between the devices and the market is simulated through aggregation and disaggregation processes aimed at optimally managing the available flexibility from large set of devices.

- *Market layer*: The core of the simulator is represented by the optimization algorithm, responsible of simulating the balancing market clearing process, designed in order to manage large optimization problems (incl. the constraints of all the networks) and the TSO-DSO interaction models.

In order to interact with the lab, two additional functions were created. The first function pauses the simulation until the laboratory updates the input file with the characteristics of the devices (e.g., active power) and their corresponding bid. When the data is updated, the first function reads and converts the data into the correct format and inserts it into the database. Then, the platform takes into account the lab devices to compute the clearing of the market and the network's state. The market clearing determines if the bids are accepted, while the physical layer computes the automatic secondary regulation and the reactive power modulation. After the processing of the physical layer, the second function writes the results to a file that is read by the lab.

## IV. VALIDATION TEST CASES

Three validation test cases were selected for laboratory evaluation: two tests relate to the Italian pilot and one test relate to the Danish pilot.

### A. TC1: Validation of Controllers used in the Italian Pilot

*1) Overview:* The Italian pilot aims to implement new features in order to promote the integration of ancillary services from DERs, following three objectives. First of all, the real-time observability of the Medium Voltage (MV) and LV sources by aggregation of information at the interconnection point between MV and High Voltage (HV) should be implemented. This information is provided from the DSO to the TSO. Another objective is voltage regulation, and in particular the development of an architecture and the implementation of hierarchical systems for the reactive power regulation by generators connected to HV and MV grid. Furthermore, power/frequency regulation with regard to generators connected to MV grid is still not available in the Italian market and should be investigated.

A main component in the Italian pilot is a Supervisory Control and Data Acquisition (SCADA) system with Distribution Management System (DMS) services, installed in the control center of the DSO. The DMS is specifically developed for this pilot. Furthermore, a Power Plant Controller (PPC) is designed to manage different generators. With these two components multiple DERs can be aggregated as a Virtual Power Plant (VPP), which simplifies the management of distributed resources, allowing them to participate in the market.

In this validation case, the SCADA/DMS and PPC will be tested in a lab environment to realistically assess which is the best scheme to provide flexibility and ancillary services from DERs to the grid. To do this, the SCADA/DMS and the PPC will be integrated in the lab using a HIL concept. The SCADA/DMS and the PPC will be present as real components.

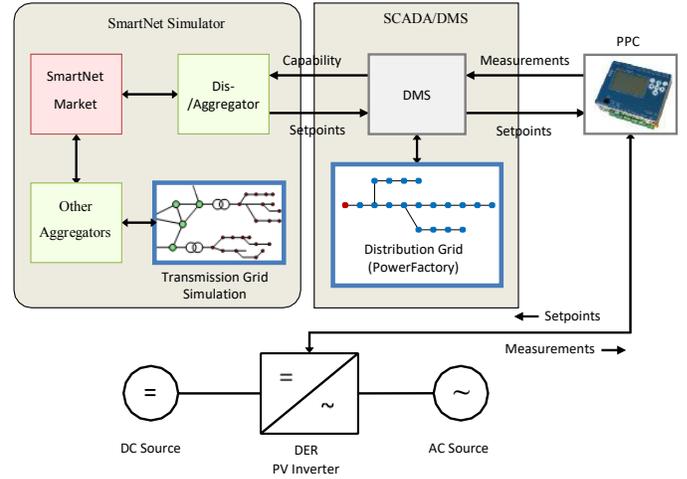

Fig. 2. Test setup of TC1: components under study are the DMS and PPC.

*2) Test Setup:* The idea with the validation case is to integrate the SCADA/DMS into the market layer using an intermediate aggregator. The role of the aggregator is to create market bids based on the capabilities offered by the SCADA/DMS. The bids are used by the simulation platform, together with the requested contribution in terms of secondary frequency regulation, in order to calculate set points, which are sent to the SCADA/DMS. Once the SCADA/DMS receives a setpoint, it is used to calculate a concrete setpoint for the PPC. An overview of this test case is shown in Fig. 2.

On the SmartNet simulator, the Italian transmission grid is simulated together with different aggregators and the SmartNet market. A hypothetical 2030 Italian scenario is used for the simulation. For the integration of the SCADA/DMS, an additional aggregator is simulated within the platform. It has the responsibility of creating bids based on the current capability sent from the DMS. The bids are then used in the market clearing and if accepted the aggregator creates setpoints for the DMS. The DMS is responsible for calculating the optimal activation setpoints of the DERs based on setpoints from the aggregator and real time measurements coming from the one or more PPCs. To do this, the DMS uses the load flow simulation of the distribution grid, running on DIgSILENT PowerFactory. Because of this, the same simulations can also be used for evaluation purposes. The DMS is also connected to a PPC, which converts the setpoints of the DMS into proper commands for the PV inverter. At the same time, measurements are also collected by the PPC from the inverter and sent to the DMS, where they are processed by the on-board distribution grid simulation.

*3) Performed Experiments and Results:* Due to the fact that the PV inverter connected to the laboratory has a nominal power of 5 kW, it was scaled-up to higher power ratings for inclusion into the SmartNet simulation. Otherwise, the DER would not have been noticeable at the transmission level at all. However, also when scaled-up, the standard Italian scenario, used in the experiment, was still too static for any noticeable changes on the DER side. Therefore, the scenario was modified and the network unbalance were increased in order to increase the participation of distributed resources. This activates more secondary regulation, which causes changes in the active power setpoint for the PV inverter. The results from this test are provided in Fig. 3.

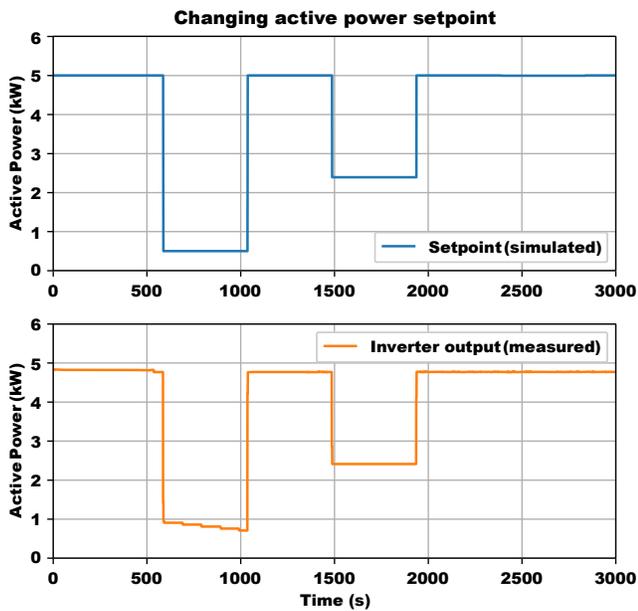

Fig. 3. Results from a simulation with an adapted scenario.

The upper plot in the figure shows the simulated setpoints sent from the aggregator in the SmartNet simulator (see Fig. 2). For comparison, the lower plot shows the measured active power output of the PV inverter. The small mismatches, especially between 500 s until shortly after 1000 s, are due to the internal control of the DMS, seen in Fig. 2.

In conclusion, the initial Italian scenario needed adaption in order to measure any interesting results from the PV inverter. In retrospect, this is also a logical outcome since the analysis done in the SmartNet simulator is focused on the transmission grid and the PV inverter is connected at the distribution grid level. Nevertheless, the tests still produced valuable results for the SmartNet project. Several issues with the DMS, the PPC, and the coupling between the components were discovered and solved during the setup of this test case. As a result, these detected issues could be avoided in the Italian pilot.

### B. TC2: Validating the Impact of ICT on the Italian Scenario

*1) Overview:* This case is a variant of the above discussed TC1, but with the addition that a communication emulator is used to analyze how a non-ideal communication network affects the interactions between the SCADA/DMS and the PPC. As part of the SmartNet project, ICT requirements for the different coordination schemes were discussed and evaluated on a theoretical basis. One of the main outcomes of this work was that modern telecommunication technologies, such as 4G or 5G, are more than capable of handling the SmartNet solutions [8]. In this validation case a dedicated communication emulator will be used to emulate exactly this kind of communication technologies [9]. Using this possibility, the intention with this validation case is to provide measurements that can be compared to theoretical analysis.

Another result of the theoretical analysis of the ICT requirements, was that communication latency will probably not have any direct effects on the performance of the coordination schemes since the update cycle is long. However, other communication network effects, such as packet loss or corrupted packages, will probably also affect the performance and thus should be taken into account [10].

*2) Test Setup:* Since this test case is an extension of TC1, the setup is very similar. Compared to the setup of TC1 in Fig. 2, the only difference is that the communication emulator is connected between the SCADA/DMS and the PPC. The emulator gives an opportunity to mimic the behavior of different types of fixed and wireless communication technologies in different radio conditions. The emulated communication link helps detect possible bottlenecks in a system and assess the impact of a communication link on the system performance. Testing components in different conditions gives better understanding about their performance margins. It is also feasible to test new technologies in the laboratory before they are deployed in the operation environment.

The communication emulator is an Ethernet switch with added functionality to emulate slower and less reliable (than Ethernet) communication paths for the selected packet flows. The flow selection is based on source and destination addresses and port numbers. The throttling and downgrading of the flow is activated by attaching a profile to the flow. A profile consists of segments and for each segment multiple parameters can be defined, such as bandwidth, delay, packet loss, etc. [9].

This means that communication profiles can be applied to all the IEC 61850 communication between the SCADA/DMS and the PPC. Since the DMS is also responsible of disaggregating the control signal for the secondary frequency regulation, it has to send a new active power setpoint to the PPC every ten seconds. Potential problems can occur if these setpoints do not arrive correctly. Thus, the communication profiles were applied to these messages.

*3) Performed Experiments and Results:* Some results from TC2 are seen in Fig. 4. A profile representing communication over a General Packet Radio Service (GPRS) network was used as a basis and on top of this a packet loss was added. For the base case, no problems were detected in the communication between the SCADA/DMS and the PPC. Once the packet loss was increased, there were cases where the system was not able to comply within the 10 s interval. This is especially shown in the lower plot of Fig. 4, where the packet loss was increased to 25 %. This is an unnatural high percentage, but was used in this case to see the effects of a package loss.

A conclusion from this test is that the implementation of the different SmartNet coordination schemes is possible even with today's ICT technologies. Based on the lab test it seems that also older technologies are able in most situations to comply to the 10 s interval. This can also be expected since ten seconds is more than enough for most ICT-based applications.

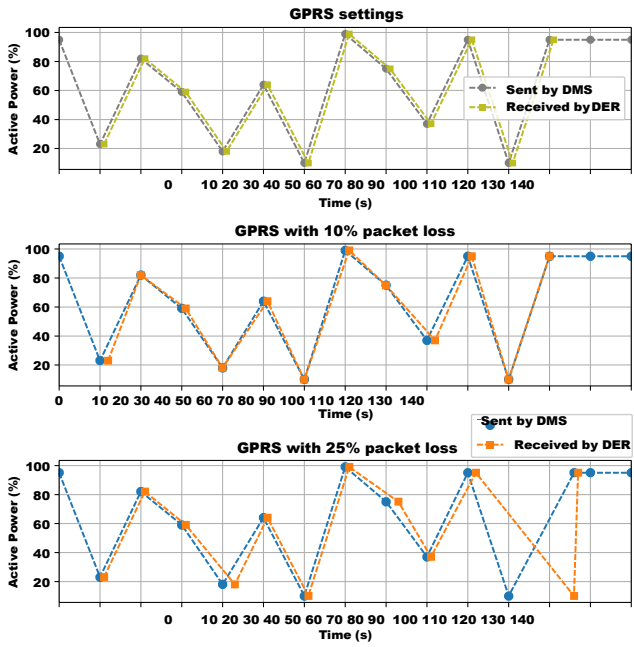

Fig. 4. Results from TC2 with an emulated GPRS network and different settings for the packet loss.

### C. TC3: Price Based Controls in Combination with SmartNet Coordination Schemes

*1) Overview:* In 2017, it was seen that 44 % of the electricity load in Denmark was covered by wind power generation. This large penetration of the stochastic wind power often leads to balancing problems. The Danish pilot aims at assessing the potential of ancillary services provision from an aggregation of Danish summer houses with swimming pools. The latter consume substantial amounts of electricity for water heating. At the same time, swimming pools have a large thermal mass. Thus, the load to heat pool water can be shifted with little consequences on the comfort of the occupants.

The Danish pilot uses a priced based control of the heater for the swimming pools [11]. Price-based controllers are different from those that were initially considered during the design of the coordination schemes. Thus, a laboratory test combining the price-based controllers with a simulation of the SmartNet coordination schemes provides an interesting test case.

*2) Test Setup:* For this test, the swimming pools at the summer houses in Denmark were connected to the laboratory in a HIL setup. Thus, the aggregator developed for the Danish pilot was also included in the setup. Furthermore, since real-time measurements of the active power consumption are available for the summer houses, the reaction of the swimming pool heaters can be monitored and used for the evaluation of the test. For the laboratory test case two summer houses were integrated. An overview of this test case is depicted in Fig. 5.

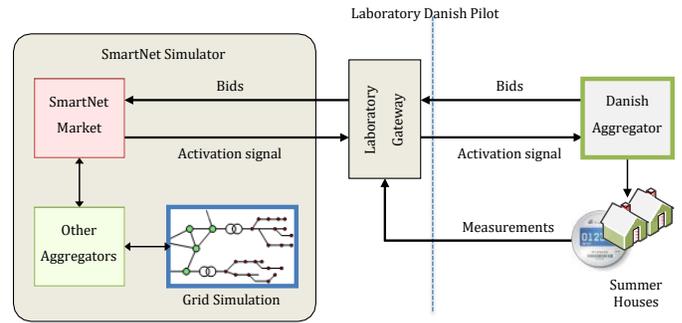

Fig. 5. Setup of TC3: integration of the price-based control with the simulator.

The setup of this test case uses the SmartNet simulator, where a scenario based on the Danish pilot is simulated. Together with the simulation of the national transmission grid, the SmartNet simulator also simulates the market and any other aggregators. Since only two summer houses are aggregated in Denmark, other virtual aggregators are also needed to be able to simulate a proper market.

In order to integrate the components from the Danish pilot, three main connections are needed. First of all, bids sent from the Danish aggregator must be forwarded and integrated into the SmartNet market simulation. Secondly, once the market is cleared, activation signals are created for all accepted bids and sent back to the aggregators. Thus, the Danish aggregator also has to be able to receive these activation signals. Thirdly, the reaction of the summer houses needs to be monitored. These three connections are seen on the right-hand side of Fig. 5. To interface the SmartNet simulator with the components running in the Danish field, a gateway installed in the SmartEST laboratory was used. The gateway was also responsible for recording the response of the summer houses.

*3) Performed Experiments and Results:* The main goal with this test is to see how well the price-based controls of the Danish pilot interacts with the SmartNet coordination schemes. They have been integrated into the market simulation but were initially not intended to be used directly with price-based controls. Therefore, the market layer of the SmartNet simulator was not developed to directly handle such situations.

The main difference of a price-based solution compared to a direct-control solution is that the actual response of the controlled system (i.e., the water heater of the swimming pools) is not known when a price signal is sent. Consequently, a control scheme designed for a direct-control solution that is used with a price-base control solution will most certainly not produce the exact same results. In Fig. 6, the results are shown from a test that was carried out on July 25, 2018. The upper plot shows the measured active power of the two summer houses. The middle plot shows the state of the pump heating the swimming pools and the lower plot shows the activated bids from the market simulation.

As seen in the figure there is a certain mismatch between activated bids and actual activations of the swimming pool heaters. This is because the summer houses have multiple criteria for activating the heaters. Therefore, the heaters may also be activated without any activation signals coming from the market. Also, the tests were run in the middle of summer, with outside temperatures of around 30° C, so there was no need to additionally heat the water.

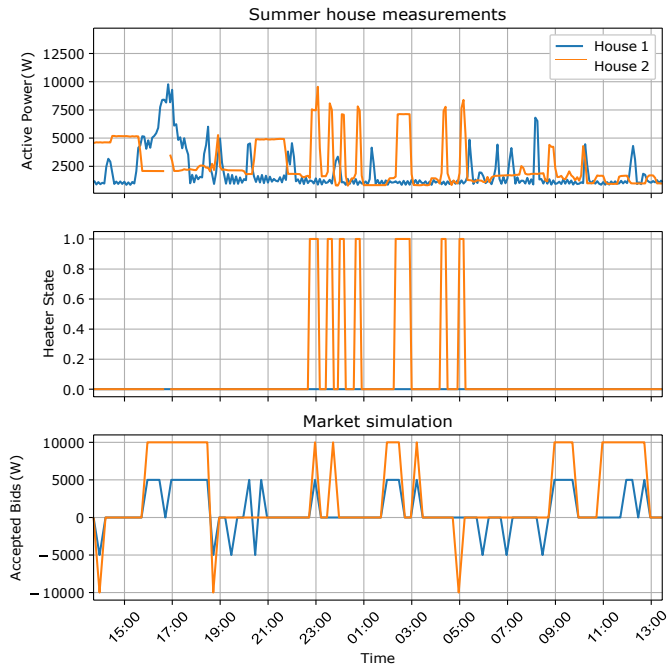

Fig. 6. Test results for integration of price-based controls with the SmartNet coordination schemes; test started 2018-07-25 at 2 pm.

The results show that the price-based control can be integrated with the SmartNet coordination schemes. However, they also show that although this is technically feasible the expected results may be something different with respect to a direct-control solution. It should also be pointed out that these tests were done with only two aggregated summer houses. By aggregating additional houses, there is statistically a higher probability for the aggregator to make accurate bids.

## V. CONCLUSION

The current increasing integration of RES into the power system is causing challenges, both for the TSO and the DSO. Using ancillary services to handle these challenges is seen as one of the key measures that needs to be integrated into future systems. Since this is a problem that can be monitored from the transmission and distribution system's sides, the coordination between TSO and DSO will be necessary. The SmartNet project compares five different TSO-DSO coordination schemes for this purpose. Besides pilot tests in Denmark, Italy, and Spain, these coordination schemes are also validated using a laboratory-based approach.

Three different validation cases have been executed: a validation of a DMS and a PPC used in the Italian pilot, the impact of communication network characteristics, and the compatibility of price-based controls with the TSO-DSO coordination schemes. These test cases were implemented using HIL setups at the SmartEST laboratory.

From the results, a number of conclusions were made. First of all, it was possible to analyze the setup of selected equipment and elaborate suggestions about their utilization in the pilots and also to analyze how the components interact with other equipment, that were not directly included in the pilots. Secondly, through the tests of the communication aspects it was shown that even current ICT technologies are capable of handling the SmartNet coordination schemes. Thirdly, the tests with the price-based control also showed that this needs special consideration within the TSO-DSO coordination schemes.

The work here also shows how laboratory tests can complement field trials. Although many aspects can be covered in field tests, there are still limitations, such as when the current regulatory framework is blocking. Another possibility is to use laboratory tests to pre-check field equipment, thereby reducing the amount of time, and often costly manual work, needed for error correction in the field trials.

ACKNOWLEDGMENT

This work has received funding from the European Commission's Horizon 2020 Program (H2020/2014-2020) under the project "SmartNet" (Grant Agreement No.691405).